\documentclass[aps,pra,letterpaper,10pt,twocolumn,superscriptaddress]{revtex4-1}
\usepackage[colorlinks=true,allcolors=blue]{hyperref}
\usepackage{amssymb}
\usepackage{amsmath}
\usepackage{graphicx}
\usepackage[caption=false]{subfig}
\usepackage{amsfonts}
\usepackage{xcolor}
\usepackage{epsfig}
\usepackage{color}
\usepackage{bm}
\usepackage{tabularx}
\usepackage{multirow}
\usepackage{mathtools}
\usepackage{xfrac}
\usepackage{blkarray}
\usepackage{bbold}
\usepackage[mathscr]{euscript}
\usepackage{autobreak}
\usepackage{comment}
\usepackage{gensymb}

\newcommand{\abs}[1]{\left|#1\right|}

\begin{document}
\title{A Coherence-Protection Scheme for Quantum Sensors Based on Ultra-Shallow Single Nitrogen-Vacancy Centers in Diamond}

\author{Anton Pershin}
\thanks{These authors contributed equally to this work.}
\affiliation{HUN-REN Wigner Research Centre for Physics, P.O.\ Box 49, H-1525 Budapest, Hungary}
\affiliation{Department of Atomic Physics, Institute of Physics, Budapest University of Technology and Economics,  M\H{u}egyetem rakpart 3., H-1111 Budapest, Hungary}

\author{Andr\'{a}s T\'{a}rk\'{a}nyi}
\thanks{These authors contributed equally to this work.}
\affiliation{Department of Physics of Complex Systems, E\"otv\"os Lor\'and University, Egyetem t\'er 1-3, H-1053 Budapest, Hungary}
\affiliation{MTA–ELTE Lend\"{u}let "Momentum" NewQubit Research Group, P\'azm\'any P\'eter, S\'et\'any 1/A, 1117 Budapest, Hungary}

\author{Vladimir Verkhovlyuk}
\thanks{These authors contributed equally to this work.}
\affiliation{HUN-REN Wigner Research Centre for Physics, P.O.\ Box 49, H-1525 Budapest, Hungary}

\author{Viktor Iv\'{a}dy}
\affiliation{Department of Physics of Complex Systems, E\"otv\"os Lor\'and University, Egyetem t\'er 1-3, H-1053 Budapest, Hungary}
\affiliation{MTA–ELTE Lend\"{u}let "Momentum" NewQubit Research Group, P\'azm\'any P\'eter, S\'et\'any 1/A, 1117 Budapest, Hungary}
\affiliation{Department of Physics, Chemistry and Biology, Link\"oping University, SE-581 83 Link\"oping, Sweden}

\author{Adam Gali}
\email{gali.adam@wigner.hun-ren.hu}
\affiliation{HUN-REN Wigner Research Centre for Physics, P.O.\ Box 49, H-1525 Budapest, Hungary}
\affiliation{Department of Atomic Physics, Institute of Physics, Budapest University of Technology and Economics,  M\H{u}egyetem rakpart 3., H-1111 Budapest, Hungary}
\affiliation{MTA–WFK Lend\"{u}let "Momentum" Semiconductor Nanostructures Research Group, P.O.\ Box 49, H-1525 Budapest, Hungary}

\date{\today}
\begin{abstract}
Recent advances in the engineering of diamond surfaces make it possible to stabilize the charge state of 7-30 nanometers deep nitrogen-vacancy (NV) quantum sensors in diamond and to remove the charge noise at the surface principally. However, it is still a challenge to simultaneously increase the action volume of the quantum sensor by placing NV centers 0.5-2 nanometers deep and to maintain their favorable spin coherence properties which are limited by the magnetic noise from the fluctuating nuclear spins of the surface termination of diamond. Here we show by means of first principles simulations that leveraging the interplay of the surface-induced strain and small constant magnetic fields, the spin coherence times of the ultra-shallow 1-nanometer deep NV center can be significantly enhanced near the spin-phonon limited regime at room temperature in $^{12}$C enriched diamonds. We demonstrate that our protocol is beneficial to $\sim$10-nanometers deep NV centers in natural diamond too where the variable coherence properties of the center to the direction of the small constant magnetic fields establish vector magnetometry at the nanoscale.    
\end{abstract}

\maketitle

%
%
\section*{Introduction}

Quantum sensors utilizing NV$^-$ centers in diamond have the potential to revolutionize nuclear magnetic resonance (NMR) spectroscopy and quantum magnetometry~\cite{doherty2013nitrogen,  balasubramanian2008nanoscale,  schirhagl_nitrogen-vacancy_2014, schmitt_submillihertz_2017, boss_quantum_2017, smits2019two, zhang_toward_2021, allert2022advances,  scheidegger2022scanning, aslam_quantum_2023}. However, their effectiveness hinges on optimizing the distance of NV$^-$ centers from surfaces while maintaining qubit state control, both crucial for enhancing the signal-to-noise ratio. The strength of NV-NMR signal diminishes with the cube of the distance to the target spins ($d$) \cite{schirhagl2014nitrogen}, necessitating the placement of NV$^-$ centers close to surfaces. However, shallow NV$^-$ centers (5-15~nm from the surface) often exhibit reduced spin-coherence times ($T_2$), which critically impacts spectral resolution~\cite{rosskopf2014investigation}. This arises from electrical and magnetic interactions with the surrounding charges and spins. Controlled interfacial band bending can mitigate electric noise and stabilize the negative charge state of NV$^-$ centers~\cite{kim2015decoherence, neethirajan2023controlled, freire2023role}; critical for very shallow NV$^-$ centers susceptible to converting to the neutral NV$^0$ state~\cite{yuan2020charge}. The challenge of residual magnetic noise remains.

A promising approach to address magnetic noise involves operating in specific magnetic fields, where the qubit levels are protected from magnetic fluctuations, also known as clock transitions~\cite{miao_universal_2020}. Achieving this requires a finite transverse zero-field splitting, which is difficult for NV$^-$ centers due to their inherent symmetry. Previous attempts to harness this technique for shallow NV$^-$ centers have yielded only limited success. Specifically, an enhancement of $T_{2}^{*}$ time for NV$^-$ centers located deeper within high-quality diamond ($\sim 15$~nm) was reported, yet the effect vanished at nanoscale~\cite{jamonneau2016competition}. Similarly, only a modest increase in $T_{2}^{*}$ time associated with a hyperfine level anti-crossing in nanopillars was observed~\cite{wang2022zero}. They further demonstrated that a bias field created by adjacent $^{13}$C nuclear spins could shift the avoided crossing to higher fields, offering a potential avenue for AC magnetometry. Concurrently, the use of spin dyads — which involve the coupled electronic spins of NV$^-$ and P1 centers — has shown promise in enhancing coherence times through analogous mechanisms~\cite{kong2020kilohertz, meriles2023quantum}. However, no strategy has convincingly demonstrated a clear advantage for ultra-shallow NV$^-$ centers.

In parallel, a distinct challenge in quantum magnetometry is the need to ascertain the complete vector of the target magnetic field, rather than merely its projection onto the NV$^-$ quantization axis. To address this issue, various strategies have been proposed, primarily involving multiple NV$^-$ centers or the coupling of electronic spins of NV$^-$ centers to surrounding nuclear spins~\cite{maertz2010vector, nizovtsev2023vector}. While these approaches have demonstrated success, they often require specific configurations of nuclear and electronic spins within the sample, which can be difficult to control experimentally. Alternative schemes have also been explored, including those leveraging optical vortex technique~\cite{chen2020calibration}, and this area of research is currently under active development.

In this study, we investigate the coherence times of ultra-shallow NV$^-$ centers, focusing on their manipulation through applied magnetic fields. Our theoretical analysis reveals that surface strain is crucial for optimizing conditions that protect qubit states from magnetic noise, allowing high coherence times even at distances as close as 1~nm from the surface. Experimentally, we observe that surface strain enhances coherence in shallow NV$^-$ centers, linked to a hyperfine level anti-crossing. These findings have important implications for quantum sensing, suggesting that minor adjustments to the external magnetic field can substantially improve the signal-to-noise ratio. Moreover, we found that the measured $T_2^*$ coherence times exhibit asymmetry due to directional residual fields, which holds promise for advancing vector magnetometry techniques.

\begin{figure*}[h!]
\includegraphics[width=2\columnwidth]{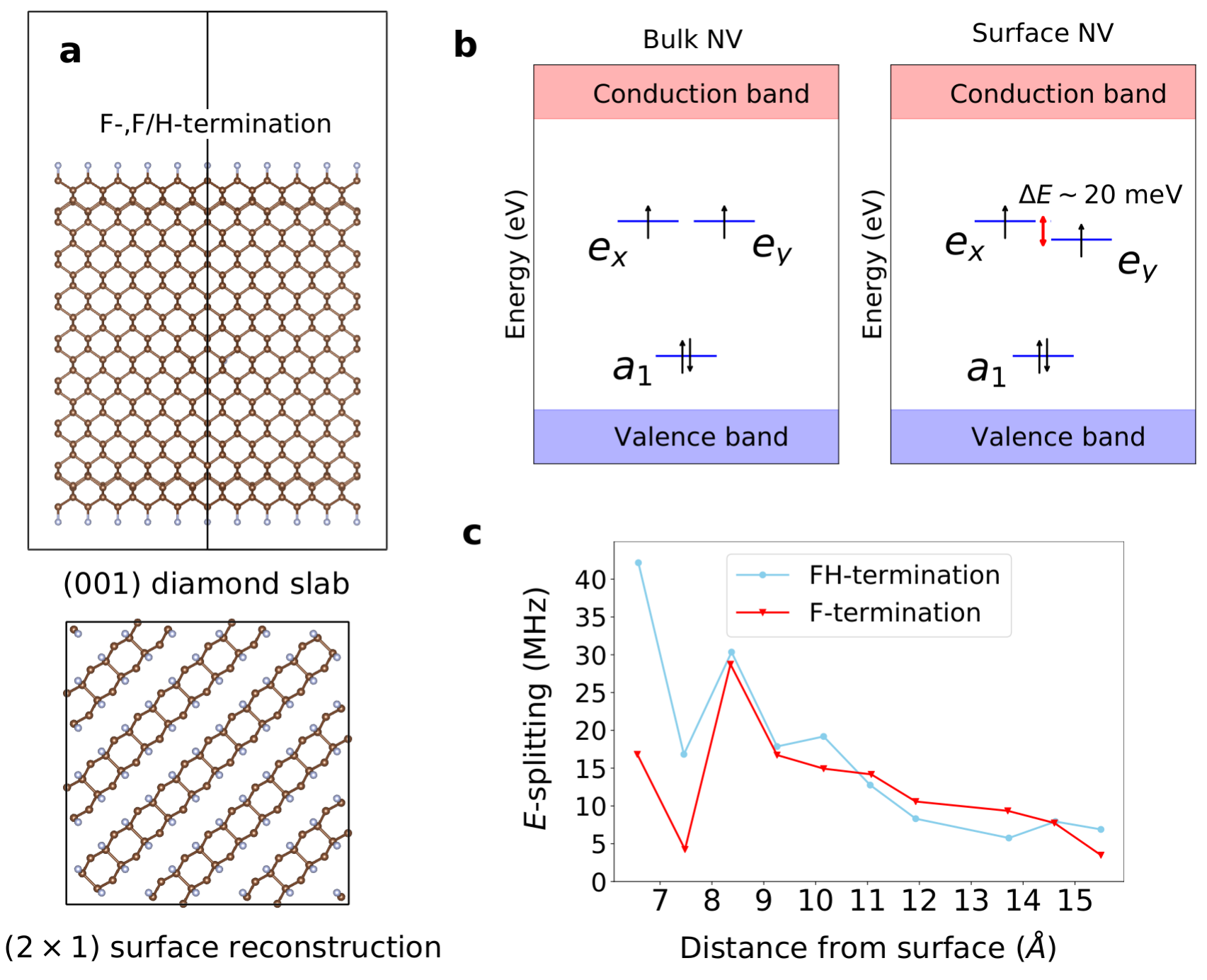}
\caption{\label{fig:NVelec}%
Electronic structure of ultrashallow NV center with electric field noiseless surface termination of diamond. (\textbf{a}) Schematic representation of (001) diamond slab used in DFT simulations. The bottom panel shows the $2 \times 1$ surface reconstruction. (\textbf{b}) Electronic structure of the NV$^-$ center in bulk (left) and close to the surface (right). (\textbf{c}) $E$-splitting as a function of NV$^-$ depth.}
\end{figure*}

\section*{Results and Discussion}

\subsection*{Suppression of magnetic noise close to the diamond surface.}

We begin our investigation with a theoretical analysis of how the surface affects electronic and magnetic properties of the NV$^-$ center. To this end, we performed DFT calculations on a large (001) slab of diamond featuring a $2\times1$ surface reconstruction pattern, see Fig.~\ref{fig:NVelec}. Two surface terminations were considered: a fluorinated surface and a mixed fluorine-hydrogen termination with a ratio of 70-30\%. This selection is based on recent advances in the modification of the diamond surface by wet chemistry~\cite{rodgers2024diamond}. Our previous study indicated that both terminations stabilize the negative charge state due to positive electron affinity~\cite{rodgers2024diamond}. In our calculations, we varied the distance of the NV$^-$ center from the surface and closely monitored changes in the electronic structure and transverse zero-field splitting (ZFS) parameters, $E$.

For NV$^-$ centers situated near the surface, we observed the lifting of $e$-orbital degeneracy, yielding an energy splitting of up to 20~meV. As depicted in Fig.~\ref{fig:NVelec}c, this symmetry-breaking resulted in a calculated $E$-value that attained a local maximum of $\sim 30$~MHz at $\sim 9$~\AA. Importantly, this effect was consistent between both surface types, leading to substantially larger $E$-values compared to those reported for the (N,H)-terminated (111) surface~\cite{korner2022influence}. It should be noted that the fluorine-terminated diamond exhibits a shallow acceptor surface state at $\sim 0.1$~eV below the conduction band minimum, which is absent in the F/H-termination case. As the NV$^-$ center approaches this surface state, no further increase in $E$-value was observed. In contrast, for the mixed termination case, the absence of surface states facilitates an even greater maximum $E$-value of $\sim 40$~MHz.

\begin{figure*}[h]
\includegraphics[width=2\columnwidth]{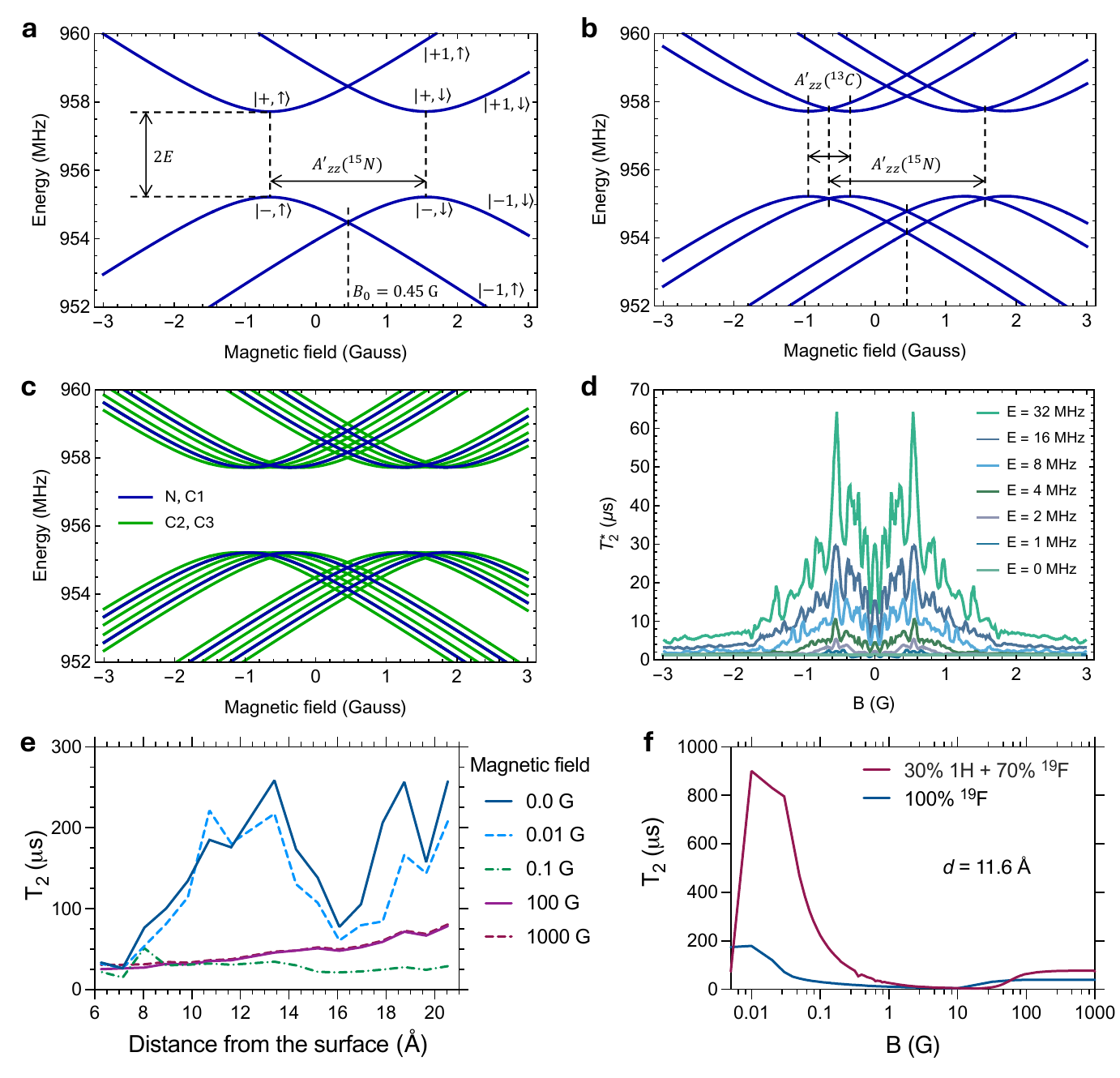}
  \caption{
  Hyperfine energy level structure in the ground state and coherence of the NV$^-$ center in diamond. \textbf{a, b, d} and \textbf{c} coupling of the electron spin to the $^{15}$N and a nearby $^{13}$C nuclear spin close to zero magnetic fields with a polar angle $\theta_0=60\degree$. The abscissa of all figures shows the total magnitude of the applied magnetic field. The shift of the curves from $B_0=0$~G is a consequence of the residual field's nonzero projection along the NV-axis. A transverse zero-field splitting of $E = 1.25$~MHz is applied to mimic the effect of the surface. {\bf a} and {\bf b} depict the case when there is $0$ and $1$ $^{13}$C nuclear spin coupled to the NV$^-$ center. {\bf c} shows multiple closely positioned splits due to two additional weakly coupled $^{13}$C nuclear spins. \textbf{d} calculated free induction decay time curves of the NV$^-$ center as a function of applied magnetic fields for different values of the $E$-splitting. The unique energy level structure induced by the nuclear spin bath gives rise to symmetric curves of complex shapes.  \textbf{b} calculated $T_{2}$ time as a function of distance from the surface for a shallow NV center at selected magnetic fields. \textbf{d} comparison between the $T_{2}$ curves computed for the NV centers in fluorinated and mixed terminations.} 
  \label{fig:LACS}
\end{figure*}
To illustrate the formation of avoided crossings in the presence of finite $E$-splitting, we further consider a typical scenario of a (001) diamond with shallowly implanted $^{15}$NV centers. The hyperfine energy level structure of the NV$^-$ center interacting with $^{15}$N and one or three $^{13}$C nuclear spins is shown in Fig.~\ref{fig:LACS}. Due to the hyperfine interaction with the nitrogen spin, at zero magnetic field, a gap opens between the pairs of states $\left\{\left| +1, \uparrow \right \rangle,\left| -1, \downarrow \right \rangle\right\}$ and $\left\{\left| +1, \downarrow \right \rangle,\left| -1, \uparrow \right \rangle\right\}$, with increased and decreased energy, respectively, where $\left| \{-1, +1 \} \right \rangle$ are the electron spin states and $\left| \{\uparrow,\downarrow\}  \right\rangle$ are the respective nuclear spin states. Applying a small magnetic field along an axis splits the electron spin eigenstates according to the parallel component $g_e\mu_BB_z = g_e\mu_B\cos{\theta_0}B_0$ of the applied magnetic field where $\theta_0$ is the polar angle between $B_0$ and the NV symmetry axis. Similarly, the coupling between the nuclear spin of a nearby carbon and the electron spin of the NV$^-$ splits the following state pairs, from higher to lower energies: $\left\{\left| +1, \uparrow, \uparrow \right \rangle, \left| -1, \downarrow, \downarrow \right \rangle\right\}$, $\left\{\left| +1, \uparrow, \downarrow \right \rangle, \left| -1, \downarrow, \uparrow \right \rangle\right\}$, $\left\{\left| +1, \downarrow, \uparrow \right \rangle, \left| -1, \uparrow, \downarrow \right \rangle\right\}$, $\left\{\left| +1, \downarrow, \downarrow \right \rangle, \left| -1, \uparrow, \uparrow \right \rangle\right\}$. The nonzero transverse zero-field splitting mixes the $\left| +1\right\rangle$ and $\left| -1\right\rangle$ states of the electron spin, leading to multiple avoided crossings (clock transitions) at magnetic field values determined by the energy-level splittings described above, more precisely by the $A_{zz}$ components of the hyperfine matrices, scaled by $A_{zz}/\cos{\theta_0}$. These clock transitions give rise to coherence-protected subspaces, where the system is immune to first-order magnetic field fluctuations. As additional weakly coupled $^{13}$C spins are introduced, several closely spaced splits and clock transitions appear, resulting in an energy level structure that incorporates the unique signatures of each nuclear spin interacting with the NV$^-$ center.


To understand the potential impact of $E$-splitting on the coherence properties of NV$^-$ centers and DC magnetic sensing, we conducted spin-dynamics simulations using $^{13}$C hyperfine and ZFS parameters from our bulk DFT calculations. The dependence of $T_{2}^{*}$ coherence time on the magnitude of transverse zero-field splitting and applied axial magnetic field can be seen in Fig.~\ref{fig:LACS}d. When the $E$-splitting reaches a value larger than the $A_{zz}$ value of the strongest coupled nuclear spin, the coherence time increases by orders of magnitude at the magnetic field values of the clock transitions, see Fig.~\ref{fig:LACS}a-c. In this regime, the larger the $E$-splitting, the longer the coherence time. Supplementary Note 2 contains additional results and an in-depth analysis of the parameter dependence of the free induction decay time. 


To further comprehend the impact of $E$-splitting on the coherence properties of NV$^-$ centers, we study the evolution of spin-echo $T_2$ time as a function of the applied magnetic field ($B_0$) with and without transverse zero-field splitting. In bulk diamond (where $E$ is zero), $T_2$ time shows a smooth increase, plateauing at high magnetic fields (c.f.\ Supplementary Figure~1a). However, $E$-splitting introduces a sharp peak at the level anti-crossing close to $B=0$~G, where higher $E$-values increase $T_2$ and extend the range of magnetic fields over which this effect persists (see Supplementary Figure~1b).

Fig.~\ref{fig:LACS}e shows the calculated $T_2$ times for the NV centers in fluorinated $^{12}$C enriched diamonds as a function of depth in various magnetic fields. As NV centers approach the surface, their interaction with surface (nuclear) spins substantially reduces $T_2$ times compared to bulk values. However, the $T_2$ values remain stable at the clock transition (c.f.\ Fig.~\ref{fig:LACS}f). Our simulations suggest that an optimal balance between high coherence time and implantation depth occurs when the NV$^-$ center is about 12~\AA\ from the surface, predicting a six-fold increase in $T_2$ time compared to high magnetic fields. Notably, mixed termination yields even longer coherence times at the clock transition, with the maximum $T_2$ time of $\sim 1$~ms, which closely approaches the value for the bulk NV$^-$ center. This enhancement is due to the differing gyromagnetic ratios of the hydrogen and fluorine nuclear spins, which decouple the two spin baths~\cite{yang_electron_2014}.

\begin{figure*}
\includegraphics[width=2\columnwidth]{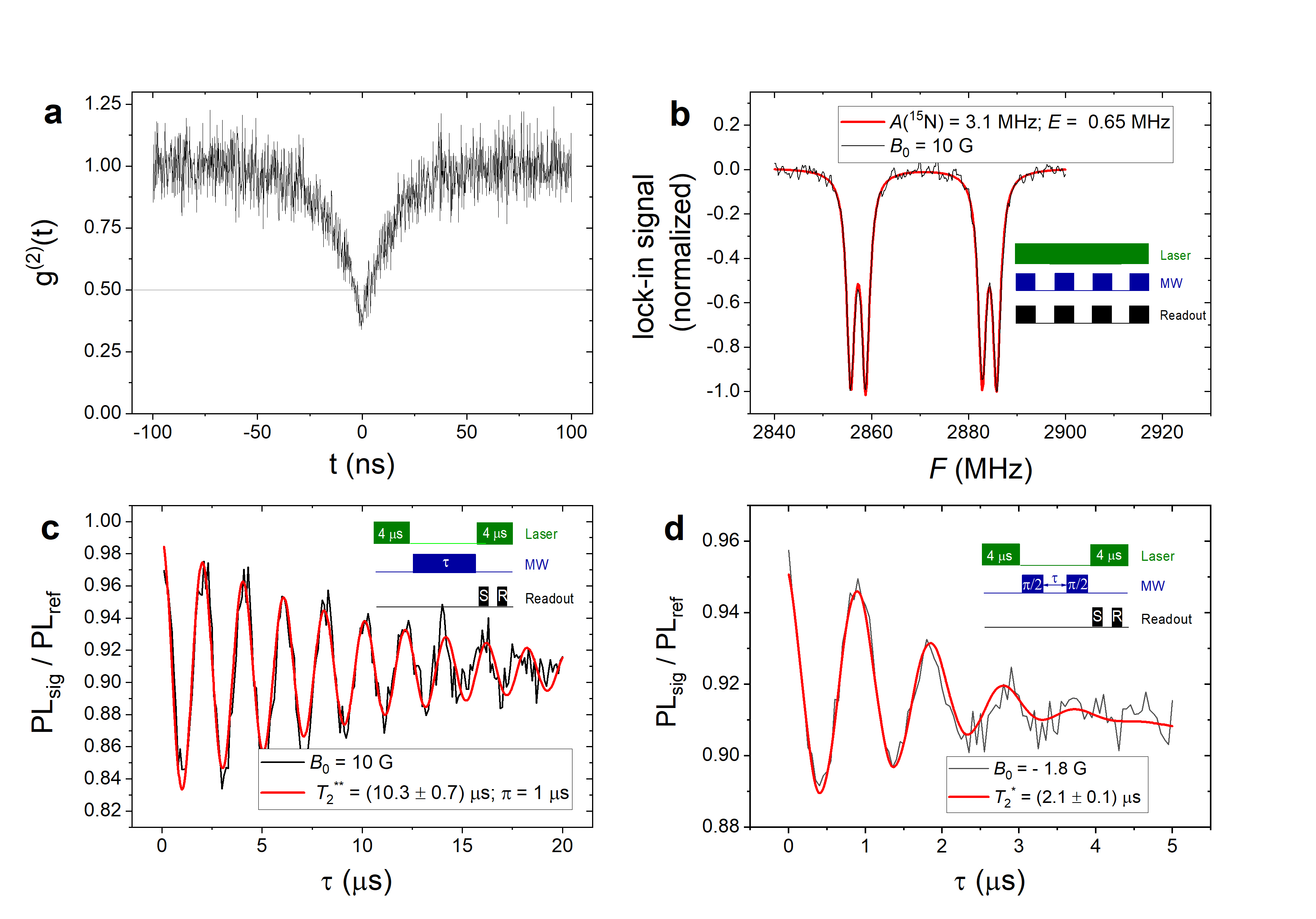}
\caption{\label{fig:NV1exp}%
The measured experimental characterization of the NV1 center. \textbf{a} $g^{(2)}$-function. \textbf{b} cw ODMR spectrum where the angle $\theta_0=61.3$ degrees between $B_0$ and the symmetry axis of the NV center. The red line shows the simulation performed using the EasySpin package. \textbf{c} The measured Rabi oscillation simulated by a cosine function with an exponential decay with time $T_2^{**}$ (red curve). \textbf{d} The measured Ramsey oscillations at $B_0=-1.8$~G simulated by a cosine function with exponential decay with time $T_2^{*}$ (red curve). The inset in \textbf{b}-(\textbf{d}) shows the measurement protocol.}

\end{figure*}

\subsection*{Near surface NV$^-$ centers in diamond nanopillars}

After discussing our first theoretical findings, we now present experimental results demonstrating how specific magnetic fields enhance coherence time in Ramsey interferometry for the currently available near-surface NV centers with natural abundance of carbon isotopes in diamond. We identified two shallow NV$^-$ centers, NV1 and NV2, within a diamond nanopillar, confirmed by the antibunching observed in the $g^{(2)}$-function measurements (c.f.\ Fig.~\ref{fig:NV1exp}a) that are located at around 8~nm deep from the surface (see also Methods). The continuous-wave optically detected magnetic resonance (cw ODMR) spectrum shows doublet peaks for these centers, with a peak splitting of 3.1~MHz attributed to hyperfine interaction with the nuclear spin of $^{15}$N~\cite{rabeau2006implantation}. Comparing simulations with experimental ODMR spectra revealed an $E$-splitting of 0.65~MHz and an angle of $\theta_0 = 61.3$ degrees between the external magnetic field (10~G) and the symmetry axis of NV1. We measured the $\pi$-pulse width to be 1~$\mu$s using a Rabi pulse sequence (c.f.\ Fig.~\ref{fig:NV1exp}c), and the typical Ramsey spectrum for NV1 together with the pulse scheme are shown in Fig.~\ref{fig:NV1exp}d.

Given the theoretical results in Fig.~\ref{fig:LACS}  
and taking into account the inclination of the NV$^-$ center in (100) diamond relative to the applied magnetic field $B_0$, the avoided crossing should be expected at $\sim 1.1$~G ($B_0^z \sim 0.55$~G), which corresponds to half of the $^{15}$N hyperfine constant. 
The ODMR spectra (Fig.~\ref{fig:NV1exp}b) show no nearby nuclear spins, such as $^{13}$C, while distant spins are obscured by a linewidth broadening of about 1~MHz. Therefore, we expect avoided crossings to appear as multiple closely positioned splits (c.f.\ Fig.~\ref{fig:LACS}d). Additionally, a magnetic environment including Earth's field acts as a bias field of $\sim 0.45$~G ($B_r$), further shifting clock transition positions relative to $B_0 = 0$~G.

\begin{figure*}
\includegraphics[width=2\columnwidth]{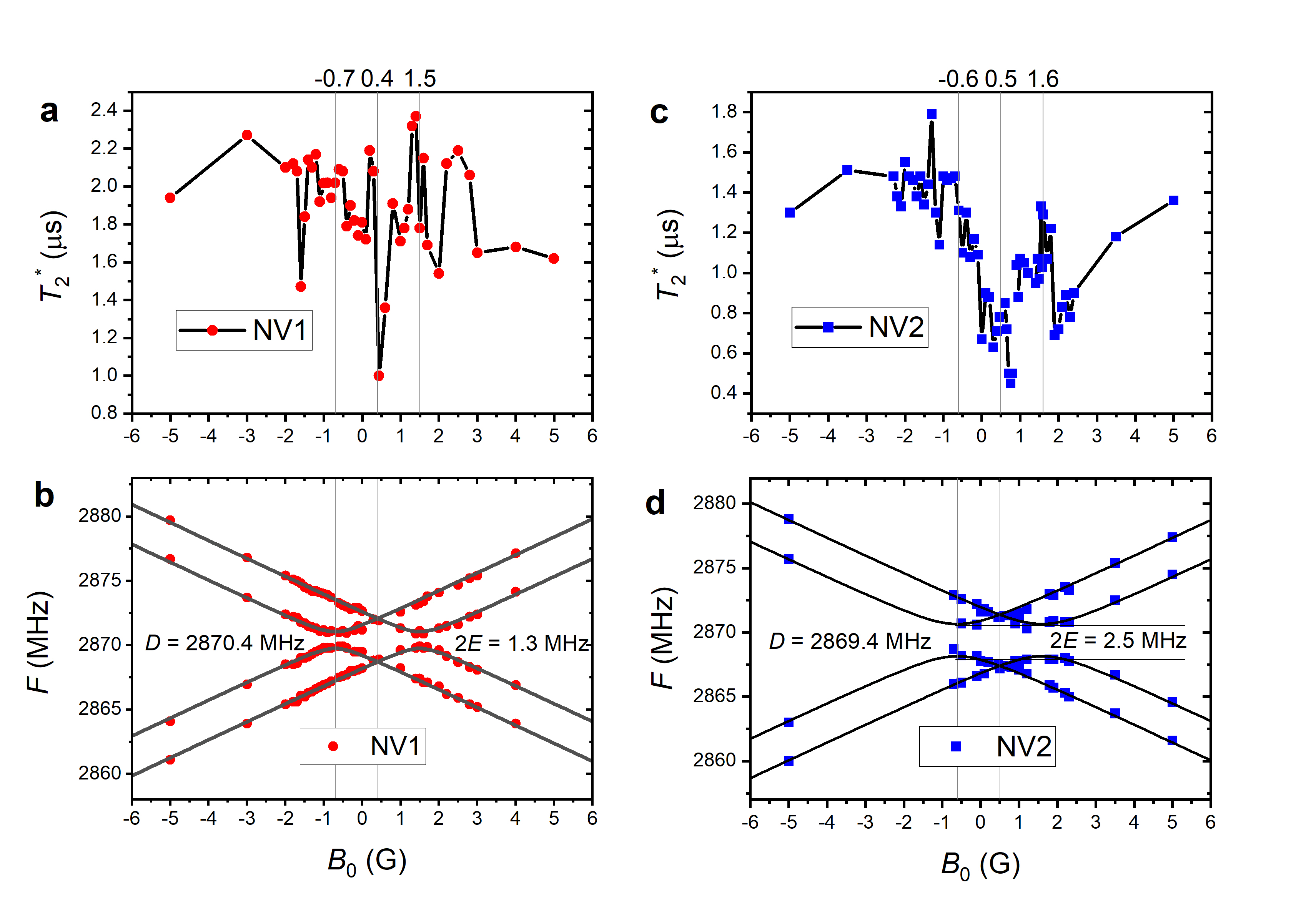}
\caption{\label{fig:NV1-NV2exp}%
Coherence of single NV$^-$ centers in nanopillars. Measured $T_{2}^{*}$ times as a function of the applied magnetic field \textbf{a} for NV1 and \textbf{c} for NV2 . The energy level structures ($F$) for \textbf{b}  NV1  and \textbf{d} NV2, derived from the positions of the maxima in the cw ODMR spectra. The solid lines represent simulations based on the indicated zero-field splitting parameters, while splits due to distant spins are omitted for clarity. $B_0$ is the applied magnetic field generated by the coil. The curves are shifted from the zero magnetic field by 0.4~G for NV1 and by 0.5~G for NV2 due to the projection of the biasing magnetic field onto the NV-axis.}
\end{figure*}

We proceed by exploring the coherence properties of the NV$^-$ centers in nanopillars in the vicinity of the avoided crossing regions. 
The time evolution of the qubit state shows the oscillations that are driven by a single frequency, corresponding to a detuning of 1 MHz from the nearest hyperfine line. 
For NV1, the measured magnetic field dependence of $T_{2}^{*}$ coherence time displays a complex behavior, particularly within the $\pm 1$~G range (see Fig.~\ref{fig:NV1-NV2exp}a).  
The expected features at the avoided crossings are not pronounced for this NV$^-$ center, which we attribute to the small $E$-value of 0.65~MHz (see Fig.~\ref{fig:NV1-NV2exp}a). 
However, several peaks and dips are discernible against a noisy constant background.

To theoretically support our measurements, we perform CCE-1 free induction decay simulations by taking into account the conditions of the measurements, including (i) the angle of the applied magnetic field and the symmetry axis of the center; (ii) the presence of the residual magnetic field of Earth; and (iii) the presence of a small transverse zero-field splitting. To account for the hyperfine shift and splitting of the clock transitions, we consider two strongly coupled nuclear spins, the $^{15}$N of the NV$^-$ center with resolvable hyperfine splitting (see Fig.~\ref{fig:NV1exp}b) and a $^{13}$C, which is adjacent to the nitrogen. To accurately model relevant dynamics, we go beyond the standard CCE-1 method by including three spins, the electron spin and the $^{15}$N and $^{13}$C nuclear spins in the central core of the system, and four spins in each subsystem of the first-order expansion of the weakly coupled bath of farther $^{13}$C nuclear spins. The upper limit on the hyperfine coupling of the spins in the bath is set to $A_{zz}^{(b)}<A_{zz}^{(C)}$, as these cannot be resolved experimentally.

Similarly to our experiments, we observe a complex structure with multiple sharp peaks and dips standing out from an overall constant magnetic field dependence of the T$_2^*$ curve, see 
Supplementary Fig.~2. We note that the features are connected to the zero-magnetic field region and they are absent at higher fields. The peaks and dips are centered around the clock transitions and $B_0 = 0$~G. Details of the curve, such as the baseline of the $T_2^*$ plot as well as heights and depths in the curve act as unique fingerprints of the arrangement of the surrounding $^{13}$C bath (beyond the hyperfine resolved close $^{13}$C), which we selected randomly in our simulations. Considering the baseline, 
we observe that the distribution of the $T_2^*$ values extends in the interval of $\sim$1~$\mu$s and $\sim$4~$\mu$s with a maximum around $\sim$2~$\mu$s. 
Our findings thus suggest that the nuclear spin bath and the induced quasi-static magnetic field fluctuation are the dominant sources of decoherence in our experiments. However, a large enhancement of the $T_2^*$ values at clock transitions for the NV1 center is neither observed in the experiment nor in the simulations, due to the comparable value of the $E$-splitting (0.65~MHz) and the hyperfine splitting of a close $^{13}$C nuclear spin, see Supplementary Notes 2 and 3.

To further comprehend the decoherence of NV$^-$ centers in nanopillars, we study yet another center named NV2. The magnetic field dependence of the ODMR spectrum for the NV2 center is illustrated in Fig.~\ref{fig:NV1-NV2exp}c. Compared to NV1, this center exhibits a substantially larger $E$ value of 1.25~MHz. Furthermore, Fig.~\ref{fig:NV1-NV2exp} shows that the $T_2^*$ time reduces 
to 1~${\mu}$s on average across all fields, except at (one of the) clock transitions, which corroborates our theoretical predictions. The combination of these two factors leads us to conclude that NV2 is located closer to the surface than NV1. 
Consequently, the peaks at the level anti-crossing become more pronounced for NV2. The observed peak splitting near $B_0 \sim 1.5$~G can be attributed to the presence of another closely coupled $^{13}$C spin, with a hyperfine coupling constant of $\sim 0.6$~MHz, which falls below the resolution of ODMR measurements. Importantly, the $T_2^*$ time at the avoided crossing, reaching 1.8~$\mu$s, is the longest observed across all studied magnetic fields within the range of $\pm 100$~G. 

As shown in Supplementary Fig.~4, our calculations for the NV2 center also confirm the maximum of the $T_2^*$ time at the avoided crossing. Furthermore, we highlight the detrimental role of $^{13}$C nuclear spins with coupling strength comparable to the transverse zero-field splitting, which are responsible for a two-fold decrease of the coherence time at the clock transition. Indeed, in the absence of $^{13}$C nuclear spins, the nitrogen nuclear spin is always present, but does not hinder coherence time enhancements. We attribute the distinct effect of nuclear spins on the $T_2^*$ time to two factors: the off-diagonal pseudo-secular Hamiltonian coupling terms arising from nonzero $A_{xz}$ and $A_{yz}$ hyperfine terms, and the increased system dimensionality, giving rise to multiple clock transitions of decreasing relevance. For completeness, we also explored various combinations of $E$-values and hyperfine constants in the Hamiltonian, as well as the impact of different hyperfine coupling terms, with the results summarized in Supplementary Notes~2 and 3.

\begin{figure}
\includegraphics[width=\columnwidth]{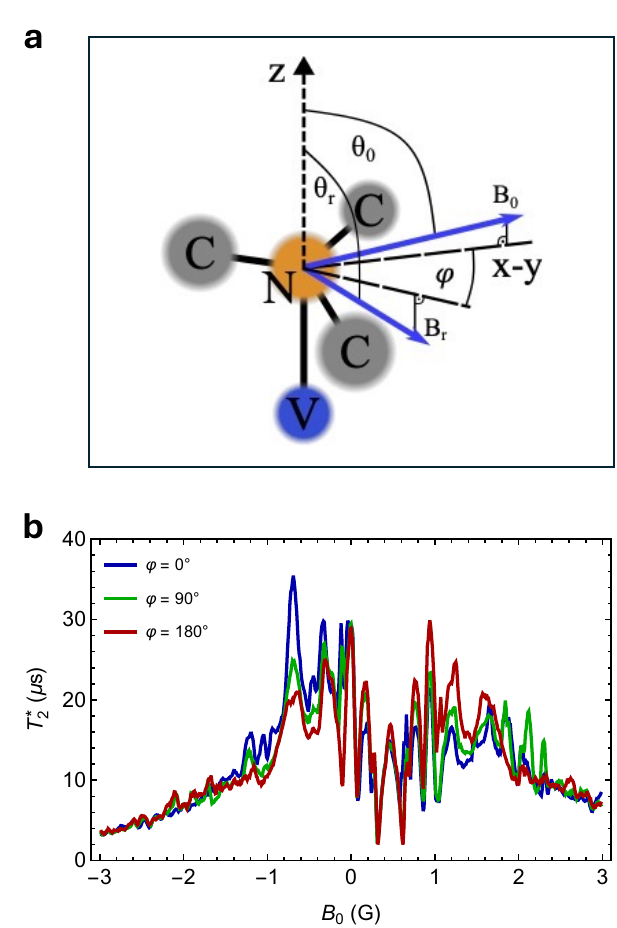}
\caption{\label{fig:NVsim}
Simulated coherence times as a function of external constant magnetic fields. \textbf{a} Illustration of magnetic field orientations relative to the NV$^-$ center and its neighboring carbon atoms. The $z$-axis corresponds to the symmetry axis of the NV center, $\mathrm{B_0}$ and $\mathrm{B_r}$ denote the applied and Earth's residual magnetic field, respectively. The dashed lines lay in the plane perpendicular to the $z$-axis. For our simulations $\theta_0=60\degree$ and $\theta_r=120\degree$. The latter is calculated from the shift of the experimental energy level spectra from $B_0=0$~G. \textbf{b} Calculated curves of free induction decay time as a function of small applied magnetic fields for the case $E=15$~MHz. $\varphi$~denotes the relative azimuthal angle between the applied and Earth's residual magnetic field.}
\end{figure}

Most strikingly, the pronounced asymmetry in the $T_2^*$ time as a function of the magnetic field $B$ is consistently observed across all our calculations for $E>1$~MHz, irrespective of the spin-bath composition, see Supplementary Figs.~5 and 6. The primary origin of this effect lies in the difference between the magnitudes of transverse magnetic field components ($\abs{B_\perp}$) at the different clock transition regions. Combined with the $A_{xx}\hat{S}_x\hat{I}_x$ and $A_{yy}\hat{S}_y\hat{I}_y$ hyperfine spin-flipping terms between the electron and nitrogen spin, these components induce mixing between eigenstates $\left|1\downarrow\right\rangle\leftrightarrow\left|0\uparrow\right\rangle$ and $\left|-1\uparrow\right\rangle\leftrightarrow\left|0\downarrow\right\rangle$ of the NV center, see Supplementary Note~3. This effect diffuses the coherence-protected regions of clock transitions by transforming the coherent spin state corresponding to the avoided crossing into a state much more sensitive to magnetic field fluctuations along the NV-axis. The frequency of this transition depends linearly on $\abs{B_\perp}$; thus, when the relative azimuthal angle of the applied magnetic field and the biased magnetic field ($\varphi$ in Fig.~\ref{fig:NVsim}a) is 180$\degree$, the field vectors are collinear and $\abs{B_\perp}$ is the same for both clock transition regions. Therefore, the curve is symmetric under this condition. In turn, the difference in $\abs{B_\perp}$ and consequently the asymmetry is maximal for the case when $\varphi$=0$^{\circ}$. 
Numerically, the strength of this effect is amplified at higher $E$-values, indicating that $T_2^*$ times can be optimized by carefully controlling the orientation of both fields. 

Surprisingly, the maximum coherence time for a given NV$^-$ center is achieved when $B_0$ is antiparallel with the target field rather than it is aligned with the NV$^-$ axis, see Fig.~\ref{fig:NVsim}b. This finding opens interesting possibilities for vector magnetometry. Specifically, following an increase in $T_2^*$ at the avoided crossing upon rotating the applied magnetic field, the maximum $T_2^*$ value would indicate the configuration when both fields are aligned, corresponding to $B_0\parallel(B_r^x,B_r^y,B_r^z)$. The accurate magnitude of $B_r^z$ can be determined from the shift of positions of avoided crossings, see Fig. \ref{fig:LACS} and Fig. \ref{fig:NV1-NV2exp}b,d. We anticipate that this straightforward approach might facilitate the determination of complete vectors of target fields, and is applicable to most of single shallow NV$^-$ centers without the need for special techniques.

%
%
\color{black}
\label{sec:summary}
In conclusion, we have theoretically identified specific magnetic fields that protect the qubit states of ultra-shallow NV$^-$ centers from magnetic noise. The precise values of these fields depend on the magnetic environment and implantation depth. Minor adjustments to the external magnetic field, typically around 0.5~G, can substantially enhance coherence for shallow NV$^-$ centers. Our experimental demonstrations on single NV$^-$ centers in a diamond nanopillar validate these predictions, showing a remarkable 2.4-fold enhancement in coherence time for the NV$^-$ center. Furthermore, our calculations indicate that using mixed fluorine/hydrogen surface terminations can further enhance coherence by reducing surface states and magnetic interaction among similar atoms, especially when the applied and bias fields are aligned. These findings provide the design rules for quantum sensors with potentially improved signal-to-noise ratios and applications in vector magnetometry.

\section*{Methods} 
\label{sec:methods}

\subsection{Theory}

The density functional theory calculations were performed by the VASP code~\cite{kresse1996efficiency, kresse1996efficient} using a plane wave basis. Projector augmented wave (PAW) potentials~\cite{blochl1994projector, kresse1999ultrasoft} were used with a cutoff energy of 400~eV. 2447-atom supercell model of diamond slab was constructed to avoid the interactions of defect with its periodic images and to apply the $\Gamma$-point sampling scheme. 
The geometry optimization and calculation of electronic properties were performed with the PBE functional. The convergence threshold for the forces was set to 0.01~eV/\AA.

Simulations of free induction decay of NV$^-$ centers were performed using the first-order generalized cluster correlation expansion (gCCE-1) method with an extended zero-order subsystem of three spins. The spin bath contains $^{13}$C nuclear spins of natural abundance within a cutoff radius of $r_{\text{bath}} = 30$~\AA.  To uncover unique features in coherence dynamics, we do not average over randomized bath configurations and instead study the dependence on specific spin distributions, see Supplementary Note 4.

The Hamiltonian used to model the system, consisting of a central electron spin and the surrounding nuclear spin bath, takes the following form:
    \begin{multline}
        \hat{H}=\hat{H}_{e-b}+\hat{H}_e+\hat{H}_b=\sum_i \mathbf{\hat{S}}^T\mathbf{A}^{(i)}\mathbf{\hat{I}}^{(i)}+\\ D\left(\hat{S}_z^2-S(S+1)/3\right)+E\left(\hat{S}_x^2-\hat{S}_y^2\right)+\\g_e\mu_B\mathbf{B}^T\mathbf{\hat{S}}-\sum_ig_N^{(i)}\mu_N^{(i)} \mathbf{B}^T\mathbf{\hat{I}}^{(i)}+\sum_{i<j}\mathbf{\hat{I}}^{T(i)}\mathbf{J}^{(ij)}\mathbf{\hat{I}}^{(j)},
    \end{multline}
where $\hat{H}_{e-b}$ is the interaction Hamiltonian describing hyperfine couplings between the NV$^-$ and its environment. $\hat{H}_e$ and $\hat{H}_b$ are the separated electron spin and bath Hamiltonians, encoding the zero-field splitting of the NV$^-$ center, the Zeeman-interaction and the magnetic dipolar interaction of nuclear spins, respectively. Note that accurate hyperfine tensors for the $^{13}$C nuclear spins were derived from first principles calculations, avoiding finite-size effects~\cite{takacs_accurate_2024}.

In order to numerically study the Hahn-echo coherence time of bulk and near surface NV centers in diamond, we implemented the second order generalized cluster correlation expansion method (gCCE-2)~\cite{yang_quantum_2008, onizhuk_probing_2021, haykal_decoherence_2022}. Coherence time calculations in bulk models were carried out by considering a $^{13}$C nuclear spin bath with natural abundance. The simulations included $\sim 1000$ first order subsystems and $\sim 2000$ second order subsystems, i.e., the convergent cut-off radii were set to $r_{\text{bath}} = 50$~\AA\  and $r_{\text{dip}} = 8$~\AA\ (see Ref.~\citenum{haykal_decoherence_2022}).

In our calculations for slab models we took into account only the spin bath of the termination due to $^{19}F$ and $^{1}H$ nuclear spins. We used $r_{\text{bath}} = 30$~\AA\  and $r_{\text{dip}} = 6$~\AA\ cut off radii that defined 300-400 first order subsystems and 3000-4000 second order subsystems depending on the distance of the NV center from the surface. In all cases we carried out averaging by taking 25 random initial state for the spin bath. The nitrogen nuclear spin of the NV center was neglected in our simulations. As a consequence, the level avoided crossing due to the transverse zero field interaction appears a zero magnetic field. In physical samples the position of the avoided crossing is shifted as demonstrated by our measurements. Accurate hyperfine tensors for the $^{13}$C nuclear spins were obtained from finite-size effect free first principles calculation using the method developed in Ref.~\citenum{takacs_accurate_2024}. 


\subsection{Experiment}

To locate a shallow single NV$^-$ center in diamond with natural abundance of carbon isotopes, we used a specialized enhanced single NV array with a pillar structure containing $^{15}$N isotopically substituted NV$^-$ centers at an average depth of 8~nm. ODMR measurements were conducted using a home-built confocal setup with a 520~nm laser (10~mW). Emission up to 700~kCs/s/center was collected through a 0.9~NA Carl Zeiss objective, filtered with a 650-nm long-pass filter, and focused into a 50:50 fiber optic coupler (Thorlabs TW670R5F1) serving as both a pinhole and beam splitter for two APD detectors (Excelitas SPCM-AQRH-44), which generated 10~ns TTL pulses for data acquisition via a Time Tagger Ultra (Swabian Instruments) module. The detectors were employed for $g^{(2)}$-function measurements (HBT interferometry). The microwave field was generated by a microwave source (WindFreak SynthNV) together with a high-power amplifier (Mini-Circuits ZHL-16W-43+), a Mini-Circuits ZASWA-2-50DRA+ switch, and a microwave stripped antenna located under the sample. The resulting microwave power was 40~dBm. 

The relaxation time $T_2^*$ was measured using the Ramsey pulse sequence. NV$^-$ centers were initially polarized to the $m_s = 0$ state with a 4~$\mu$s optical pulse. A microwave pulse train $\pi/2 \longrightarrow \tau \longrightarrow \pi/2$ was applied, followed by a second optical pulse with variable duration $\tau$ to readout the spin state. A 3~$\mu$s pulse gated the timer tagger input at the start of the second optical pulse, and an identical pulse normalized the first. A similar protocol was used for Rabi measurements, replacing the microwave sequence with a single $\tau$-pulse while scanning the pulse duration value $\tau$. Pulse sequences were managed by Pulse Streamer 8/2 (Swabian Instruments), with all ODMR signals recorded at room temperature (297~K). An external magnetic field was applied perpendicularly to the diamond surface using a Helmholtz coil, producing a homogeneous magnetic field of $\pm100$~G, controlled by a high-precision power supply (Keithley 2280S-32-6).

\section*{Data Availability}
The data that support the findings of this study are available from the corresponding author upon reasonable request.

\section*{Code Availability}
The codes that were used in this study are available upon request to the corresponding author.

\bibliography{sample}

\begin{acknowledgments}
A.G.\ and V.I.\ acknowledge the support from the National Research, 
Development and Innovation Office of Hungary (NKFIH) within the Quantum Information National Laboratory of Hungary (Grant No.\ 2022-2.1.1-NL-2022-00004). A.G.\ acknowledges the support from the European Commission within Horizon Europe projects QuMicro and SPINUS (Grant Nos.\ 101046911 and 101135699) and the QuantERA II project MAESTRO (NKFIH Grant No.\ 2019-2.1.7-ERA-NET-2022-00045).
V.I.\ acknowledges NKFIH Grant No.\ FK145395. This project is funded by the European Commission within Horizon Europe projects (Grant Nos.\ 101156088 and 101129663). The spin dynamics computations were enabled by resources provided by the National Academic Infrastructure for Supercomputing in Sweden (NAISS) at the Swedish National Infrastructure for Computing (SNIC) at Tetralith, partially funded by the Swedish Research Council through grant agreement No.\ 2022-06725. First-principles calculations were also performed using the KIF\"U high-performance computation units. A.P.\ acknowledges the financial support of J\'anos Bolyai Research Fellowship of the Hungarian Academy of Sciences.
\end{acknowledgments}

\section*{Funding}
Open access funding provided by HUN-REN Wigner Research Centre for Physics.

\section*{Contributions}
A.P.\ carried out the DFT calculations under the supervision of A.G. A.T.\ carried out the spin dynamics simulations under the supervision of V.I. V.V.\ carried out the experiments. All authors discussed the results. A.P.\ wrote the manuscript with the contribution of all authors. A.G.\ conceived and led the entire scientific project.

\section*{Corresponding author}
Correspondence to Adam Gali (gali.adam@wigner.hun-ren.hu).

\section*{Competing interests}
The authors declare that there are no competing interests.

\end{document}